\documentclass[sigconf]{acmart}
\AtBeginDocument{%
  \providecommand\BibTeX{{%
    \normalfont B\kern-0.5em{\scshape i\kern-0.25em b}\kern-0.8em\TeX}}}

\usepackage{}
\copyrightyear{2023} 
\acmYear{2023} 
\setcopyright{rightsretained} 
\acmConference[EASE '23]{Proceedings of the International Conference on Evaluation and Assessment in Software Engineering}{June 14--16, 2023}{Oulu, Finland}
\acmBooktitle{Proceedings of the International Conference on Evaluation and Assessment in Software Engineering (EASE '23), June 14--16, 2023, Oulu, Finland}\acmDOI{10.1145/3593434.3593449}
\acmISBN{979-8-4007-0044-6/23/06}

\usepackage{xcolor}
\usepackage{tabularx}

\usepackage{array} 
\usepackage{boldline} 
\usepackage{booktabs} 
\usepackage{makecell} 

\begin{document}

\newcommand{\new}[1]{{\textcolor{black}{#1}}}
\newcommand{\ok}[1]{{\textcolor{black}{#1}}}

\title{Improving the Reporting of Threats to Construct Validity}

\author{Dag I.K. Sj{\o}berg}
\orcid{0000-0002-4941-7240}
\affiliation{%
  \institution{University of Oslo}
  \country{Norway}}
  \email{dagsj@ifi.uio.no}

\author{Gunnar R. Bergersen}
\affiliation{%
  \institution{University of Oslo}
  \country{Norway}}
\email{gunnab@ifi.uio.no}

\renewcommand{\shortauthors}{Sjøberg and Bergersen}

\begin{abstract}
\textbf{Background}: Construct validity 
concerns the use of indicators to measure a concept that is not directly measurable.
\textbf{Aim}: This study intends to identify, categorize, 
assess and quantify 
discussions of threats to construct validity in empirical software engineering literature and use the findings to suggest ways to improve the reporting of construct validity issues. 
 \textbf{Method}: 
 We analyzed~83 articles that report human-centric experiments published in five top-tier software engineering journals from~2015 to 2019. 
 The articles' text concerning
 threats to construct validity was divided into segments
 (the unit of analysis) based on predefined categories. The segments were then evaluated regarding whether they clearly discussed a \textit{threat} and a \textit{construct}.
  \textbf{Results}: Three-fifths of the segments were associated with topics not related to construct validity. Two-thirds of the articles discussed construct validity without using the definition of construct validity given in the article. The threats were clearly described in more than four-fifths of the segments, but the construct in question was clearly described in only two-thirds of the segments. The construct was unclear when the discussion was not related to construct validity but to other types of validity.
 \textbf{Conclusions}: The results show potential for improving the understanding of construct validity in software engineering. 
Recommendations addressing the identified weaknesses are given to improve the awareness and reporting of CV.
\end{abstract}

\ccsdesc[500]{General and reference~General literature}

\keywords{measurement, research quality, empirical research}



\maketitle

\section{Introduction}
\label{sec:introduction}

At least in software engineering, research is intended to improve society. 
Researchers would like their research to have an impact.
One might like to demonstrate that something proposed or enhanced 
is better than some alternative.
%
%
%
In some cases, one could demonstrate improvement according to relevant success criteria using a validated measurement instrument. 
For example, less time spent could be measured using a watch. Less energy consumption could be measured using an electronic device. 

However, many success criteria and other relevant aspects are 
often elusive and
not directly measurable, for example, quality aspects of a software system, such as functional suitability, reliability, usability, maintainability and security. 
Such concepts cannot be measured directly; we need a set of indicators to measure them. 
A concept that is not directly measurable but is represented by indicators at the operational level to make it measurable is called a construct. 
The validity of a construct, 
which is called 
construct validity (CV), is defined by how adequate a concept definition is and how well the indicators represent the concept~\cite{sjobergconstruct, cook1979design, cronbach1955construct, ralph2018construct}. 

Several general guidelines for conducting and reporting empirical studies in software engineering have been published \cite{shaw2003writing, sjoberg2007future, jedlitschka2008reporting, runeson2009guidelines, kitchenham2008evaluating}. 
This paper is focused on managing and reporting threats to CV in particular. We identify various weaknesses and mistakes concerning CV and propose means for rectifying them. We address the research questions as follows:
\begin{itemize}
    \item RQ1: How are threats to CV reported in empirical software engineering?
    \begin{itemize}
        \item RQ1.1: What are the topics reported in discussions about threats to CV?
        \item RQ1.2: How is CV defined, and how is the definition reflected in the discussion on threats to CV?
	\item RQ1.3: Which types of threats to CV are reported?
	\item RQ1.4: Are the threats and constructs clearly described in discussions about threats to CV?
    \end{itemize}
    \item RQ2: How can the reporting of threats to CV be improved in light of the answers to RQ1?
\end{itemize}

An earlier study investigated CV in a set of 83 articles that report experiments published in five top-tier software engineering journals ~\cite{sjobergconstruct}. 
In this paper, we analyze the same articles regarding additional aspects of CV. 
We also quantify the articles on aspects already reported in ~\cite{sjobergconstruct}, where the unit of analysis was either the whole article or the sections on threats to CV. In this paper, the unit of analysis is the segment, which can be paragraphs, sentences or sequences of words within a sentence. 
This finer-grained analysis provided deeper insights into and further quantification of CV issues.


The remainder of this paper is organized as follows. Section 2 describes the research method, including data collection and analysis. Section 3 reports and discusses the results. Section 4 discusses limitations of the study. Section 5 concludes.

\section{Research Method}
\label{sec:method}

\subsection{Data Collection}

This investigation uses the same set of 83 articles that provide a discussion of CV and were analyzed in a study reported in~\cite{sjobergconstruct}.\footnote{References to the 83 articles can be found in \cite{sjobergconstruct}.}  
The articles report human-centric experiments~\cite{kitchenham2012trends} published from 2015 to 2019 in five journals with a high impact factor: \textit{IEEE Transactions on Software Engineering} (TSE), 
\textit{Empirical Software Engineering} (EMSE), 
\textit{Information and Software Technology} (IST), 
\textit{ACM Transactions on Software Engineering and Methodology} (TOSEM), and 
\textit{Journal of Systems and Software} (JSS).

\subsection{Data Analysis}

Both authors simultaneously read and split the subsections or paragraphs on “threats to construct validity” into text segments consisting of sentences or sequences of words that describe a particular topic. We then categorized and analyzed these segments as described below. We worked until we achieved consensus on the segmentation, categorization and other analyses. We used the variables as follows in our analysis:

\begin{itemize}
\item \textit{Topics}. Based on findings reported in~\cite{sjobergconstruct}, we divide the segments into the following topics:

\begin{itemize}

\item \textit{Definition of CV}. When a definition was discovered in a segment, it was categorized into one of the following predefined options based on the categorization used in~\cite{sjobergconstruct}. CV is concerned with

\begin{itemize}

\item how well a concept is defined and how well a set of indicators represent the concept, cf. the definition given in Section 1,
\item the relationship between theory and observation,
\item the relationship between the goals of an experiment and its results, and
\item the extent to which the variables measure what they intend to measure.
\item A category "miscellaneous" includes definitions of CV that we find difficult to consider plausible. Two examples are: CV concerns how an experiment is \textit{constructed}, and CV concerns how well the treatment reflects the cause and outcome of the effect in an experiment.

\end{itemize}

\item \textit{Imported construct}. A segment that describes a construct that is imported from established theories, standards or frameworks, or from articles that document the construct.

\item \textit{Threats to CV}. There are three types of threats to CV:

\begin{itemize}

\item \textit{Inadequate definition of concept}. It means that the definition of the concept is unclear or ambiguous, or differs from an established understanding~\cite{cook1979design}. 
\item \textit{Underrepresentation}. It means that the set of indicators used to represent a concept is too narrow to include all important aspects or dimensions of the concept~\cite{messick1989validity}.
\item \textit{Construct-representations bias}. Indicators used to represent a concept are biased if they systematically misrepresent the concept; for example, if only one method is used to collect the data, which is called mono-method bias~\cite{cook1979design}. A representation is also biased if an indicator is outside the scope of the concept; that is, the indicator does not represent the concept.

\end{itemize}

\item \textit{Internal validity}. In a study that infers an effect \textit{B} from a cause \textit{A}, a threat to internal validity is that there could be other causes than \textit{A} for the effect \textit{B}~\cite{shadish2002experimental}.

\item \textit{External validity}. It concerns whether the findings of a study are generalizable beyond the immediate study~\cite{yin2003case}.

\item \textit{Statistical conclusion validity}. It concerns statistical analysis of the variables included in cause-and-effect studies.

\item \textit{Fundamental study challenges}. A study has fundamental problems if, for example, participants cheat, do not have sufficient time to fulfil tasks or do not understand the material given (text may be incomprehensible due to translation errors).

\item \textit{Unknown}. When we cannot categorize a segment into a definition, import or any kind of threat, we describe the topic as unknown.
\end{itemize}

\item \textit{Consistent use of CV definition}. This binary variable indicates whether the description of threats to CV adheres to or reflects the definition of CV given in the same article. For example, if CV is defined as the relationship between theory and observation, is there a reference to a theory or theoretical level, or if the definition is whether a variable measures what it intends to measure, is there a description of the intention?

\item \textit{Clarity of threat}. This variable indicates how clearly a threat in a segment is described, with the possible values: unclear, somewhat clear or clear.

\item \textit{Clarity of construct}. This variable indicates how clearly a construct included in a threat or another statement is described, with the possible values: unclear, somewhat clear or clear.

\end{itemize}

\section{Results and Discussion}
\label{sec:results}

The subsections below answer research questions RQ1.1 to RQ1.4, respectively, in the context of the sample of 83 journal articles. The last paragraph in each subsection is a response to research question RQ2.

\subsection{Topics in Discussions on Threats to CV}

We identified 
\ok{310} 
segments in the articles. 
Figure~1 shows the proportion of topics in the collected segments. The figure also shows the proportion of text (number of words) used to describe the various topics. 

We consider it relevant to include the first three topics described in Section 2 (\textit{definition of CV}, \textit{imported constructs} and, of course, \textit{CV}) in a section on “Threats to construct validity”. 
These three topics
constitute approximately two-fifths of the total segments, meaning three-fifths include topics not relevant to CV. 
The essential topic "threats to CV" forms 27\% of the segments but is included in 60\% (50) of the articles. 
Still, this means that 40\% of the articles do not discuss threats to CV.

The most frequent topic in the segments is internal validity, which is discussed in 61\% (51) of the articles. The aspects of internal validity include issues such as hypothesis guessing (the participants in a study may guess the purpose of the study), evaluation apprehension \cite{rosenberg1969conditions} (the participants may be worried that their potentially poor performance may become visible to others), and experimenter expectancies \cite{cook1979design} (the researcher may intentionally or unintentionally influence the behaviour of the participants). Such issues may impact cause-effect relationships observed in a study and thus affect internal validity, but they do not influence how a concept is operationalized in terms of indicators and thus do not affect CV.


The third-largest category is the “fundamental study challenges”, which includes a range of issues as described in Section~2. 
We believe 
that the authors of these articles included these issues under threats to CV somewhat arbitrarily in the absence of a common, appropriate predefined validity type. For example, it is difficult to consider cheating and allocating too short a time for a study as related to CV.

Most of the segments in the category "statistical conclusion validity" discuss measurement scales. For example, it is suggested that a finer-grained scale might have been better than a binary scale, or the use of a Likert scale is questioned in a given study.

\begin{figure}[t]
  \centering
  \includegraphics[width=\linewidth]{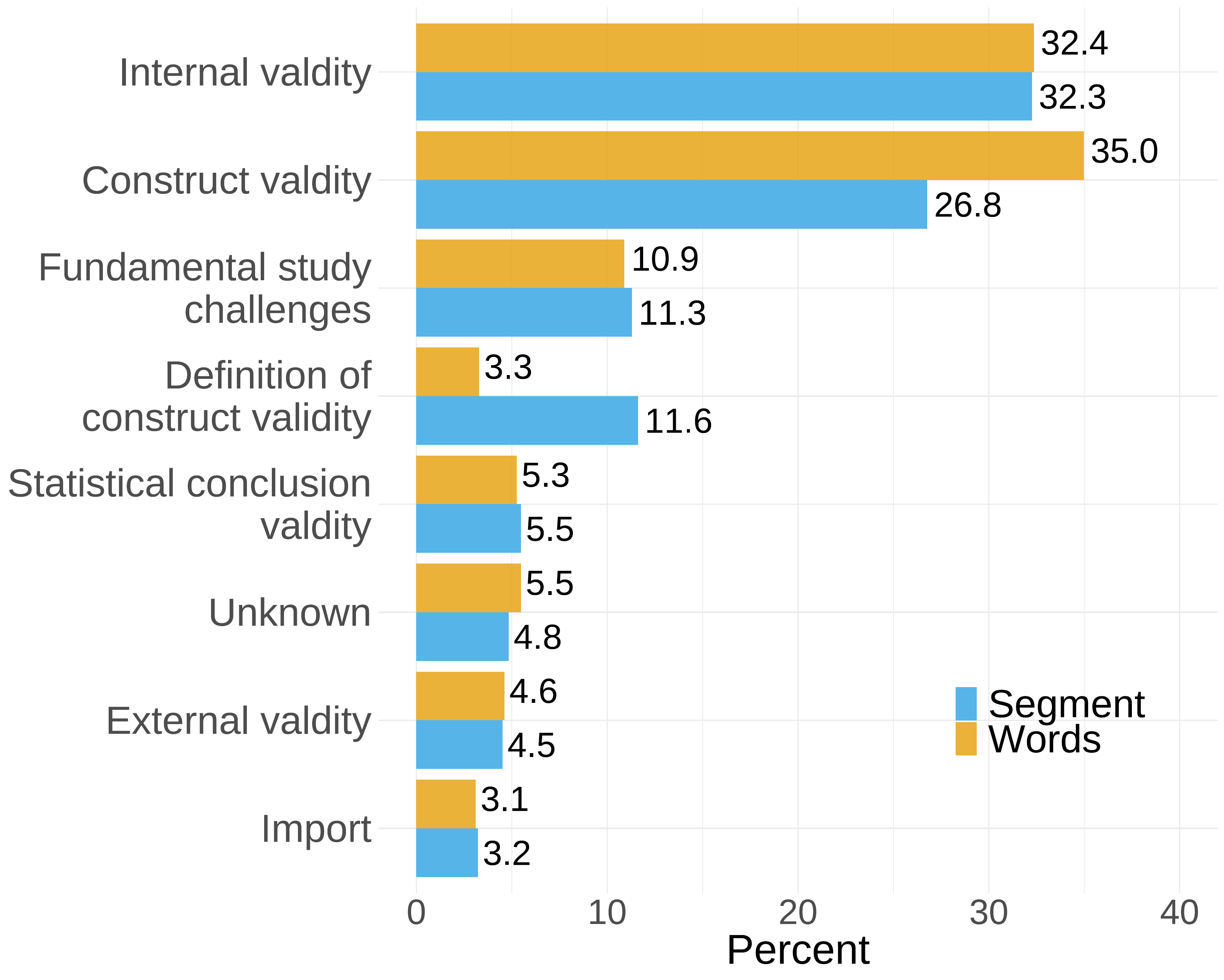}
  \caption{Topics in discussions on threats to CV}
\end{figure}

The segments categorized as external validity typically discuss generalization to other populations of people than the sample of participants or generalizations of the study material (tasks, systems, tools, etc.). Note that when populations of subjects or study material are not concepts or constructs included in research questions or hypotheses, we consider them related to external validity. If they are included as moderators in a cause-effect relationship, the operationalization of the populations would have been included in a discussion of CV. A further discussion of the relationship between external validity and CV can be found in \cite{shadish2002experimental}.

We could not categorize approximately 5\% of the segments; no threat was indicated, nor did they describe definitions of CV or imported constructs. 

Given that most of the text of sections or paragraphs entitled "threats to CV" in the analyzed articles are not specifically relevant to CV, it is plausible to recommend that authors not include threats 
other than those to CV
in such sections and paragraphs.

\subsection{Definition and Use of CV}



\begin{table}[tbp]
\caption{Number of articles with inconsistency between definition and use of CV}
  \centering
  \begin{tabularx}{\linewidth}{
    >{\hangindent=0.4em\hangafter=1\raggedright\arraybackslash}X 
    >{\raggedleft\arraybackslash}p{0.03\linewidth} 
    >{\raggedleft\arraybackslash}p{0.03\linewidth} 
    >{\raggedleft\arraybackslash}p{0.03\linewidth} 
  }
    \hlineB{2}
    \textbf{Definition} & \multicolumn{1}{c}{\textbf{Articles}} & \multicolumn{1}{c}{\makecell{\textbf{Incon-}\\\textbf{sistency}}} & \multicolumn{1}{c}{\textbf{\%}} \\
    \hlineB{2}
Relationship between theory and observation & 15 & 14 & 93 \\
 Measure of intention & 4 & 3 & 75 \\
 Relationship between experiment goals and results & 6 & 4 & 67 \\
Miscellaneous definitions & 8 & 3 & 38 \\
Represents the concepts of study & 3 & 0 & 0 \\
    \hline
Total & 36 & 24 & 67 \\
    \hline
  \end{tabularx}
  \label{tab:your_label}
\end{table}
 
For segments that included a definition or explication of CV, we investigated whether the use of the term CV in the subsequent segments was consistent with the 
stated
definition. 
Table~1 shows the definitions of CV and whether they are used consistently. 
Two-thirds of the articles that introduce a definition or explication of CV do not adhere to it themselves. 
We believe that the main reason is that the concept of CV and many of its definitions are challenging to understand. 
One indication that CV is a difficult concept is the variety of definitions. 
The only definition with no inconsistencies 
is when CV is defined as a set of indicators representing the concept.

The majority of the articles (57\%) did not include a definition of CV. 
A question is whether one needs to include a definition of CV in an empirical paper. 
In the long run, CV should be sufficiently understood 
to avoid the need for a definition to be included in papers that report threats to CV. 
However, as shown in Table~1, as long as several definitions are in use and most of the analyzed articles do not apply their own definition, we have not yet reached the level of common consensus and understanding in the community.

Our recommendation may seem 
obvious, but in light of the results reported above, it may still be helpful: If a definition is introduced in a paper, it should be followed. 

\begin{figure}[b]
  \centering
  \includegraphics[width=\linewidth]{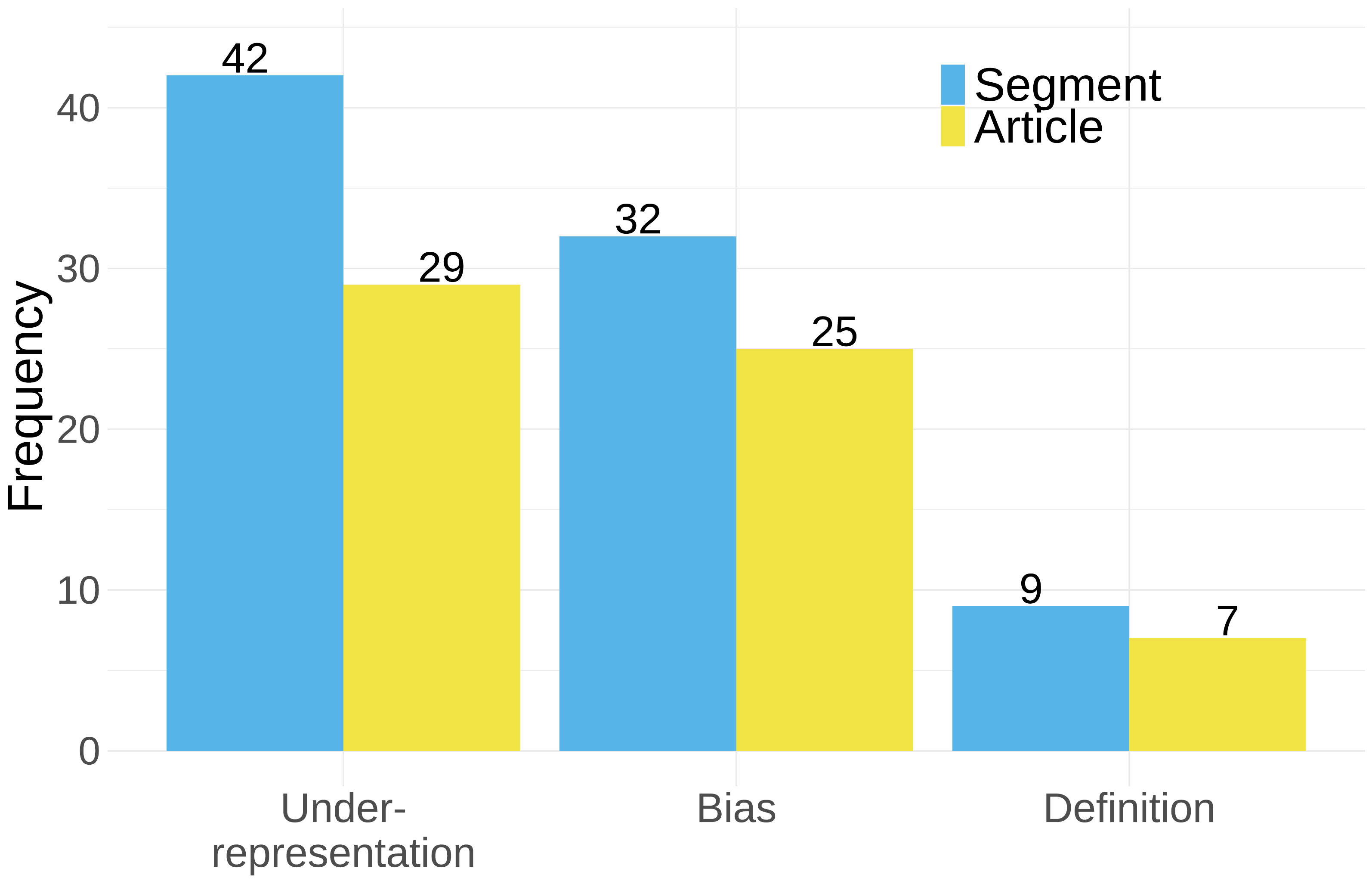}
  \caption{Types of threats to CV}
\end{figure}

\subsection{Types of CV Threats}

Figure~2 shows the distribution in the analyzed set of articles of the major types of CV threats described in Section~2. 
Most discussions concern to what extent a set of indicators underrepresent the concept in question. 
Several 
articles
explicitly discuss the threat of \textit{mono-operation}, where only one indicator is used. 
Others state that they have mitigated or even removed the threat of underrepresentation because several indicators are used. 

The discussions of bias in the representation of a concept primarily focus on the use of only one method for data collection (\textit{mono-method}) and the challenge of using subjective data. For example, there is a threat of bias if all data is collected using self-reporting. Generally, one would prefer objective to subjective data. However, one might intentionally collect subjective data, for example, if the purpose is to investigate a person’s opinions or feelings.

Note that bias in CV is not the same as bias in internal validity. 
In CV, it concerns the 
systematic (empirical) misrepresentation of a single concept.
In internal validity, it concerns whether there
are systematic problems (errors) that prevent one from making cause-effect inferences from a study, for example, because treatment and control groups are not equal in some aspect, which may unintentionally skew the study outcome. Additionally, note that "bias" in the internal validity sense may involve multiple constructs.
A high degree of CV would typically be a premise for a high degree of internal validity; bias in CV may \textit{lead to} bias in cause-effect inferences. 
However, one may still have internal validity problems despite no noteworthy problems with bias for the involved constructs.

The least discussed threat is \textit{inadequate definition of concept}. One example is an article that identifies the challenge that a tool detects a specific code smell that was not considered a smell by developers; the operational definition implemented in the tool did not comply with the developers’ understanding.

Given that relatively few articles discuss threats regarding how the operationalized concepts are defined, we recommend that authors also consider and possibly report such kinds of CV threats, in addition to underrepresentation and bias.

\begin{figure}[t]
  \centering
  \includegraphics[width=\linewidth]{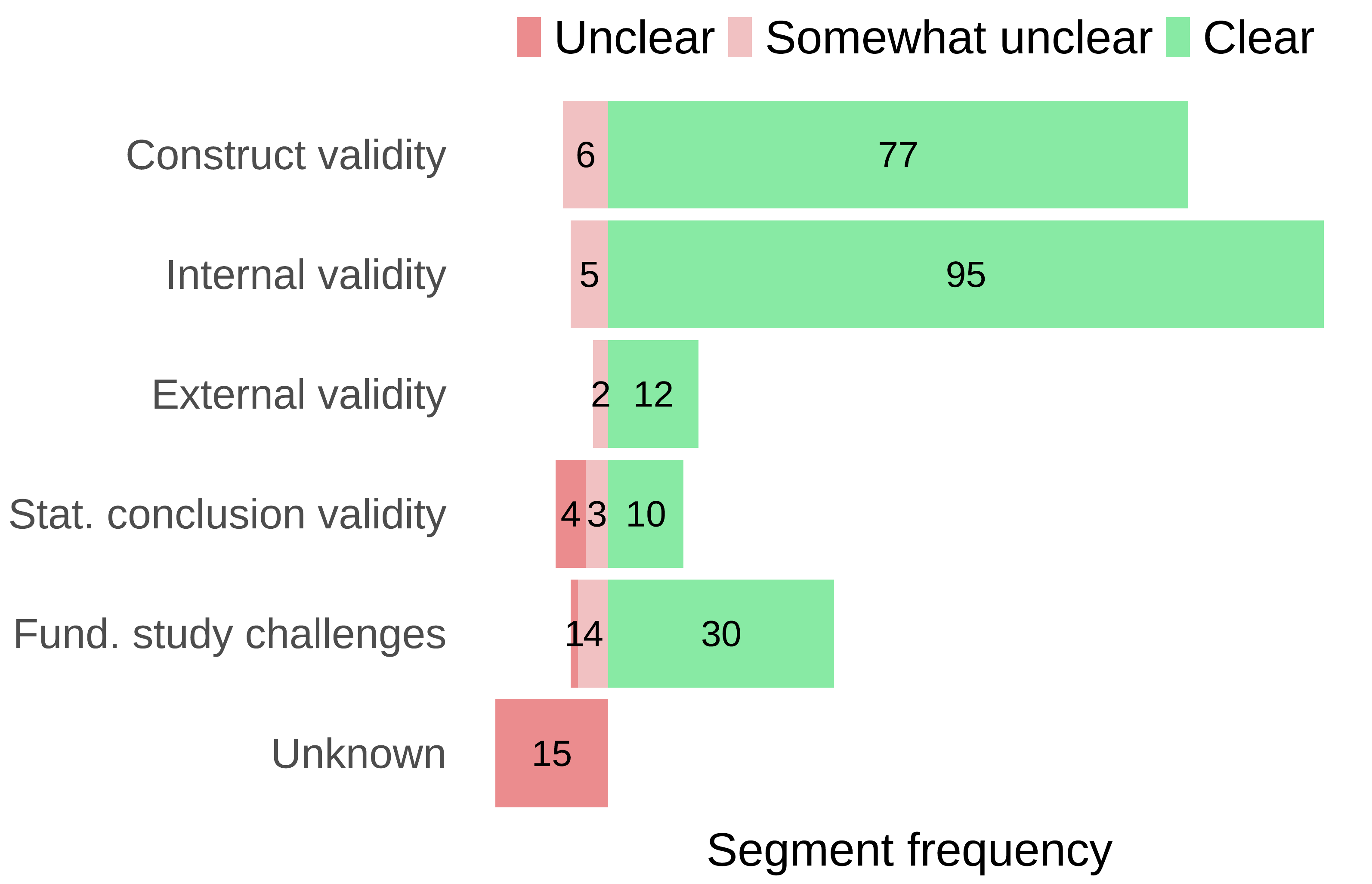}
  \caption{Clarity of threats in segments by topic}
\end{figure}

\subsection{Clarity of Threats and Constructs}

One would expect that a subsection on “threats to construct validity” would clearly describe which threats included which constructs. 
However, \ok{four-fifths of the articles have at least one ambiguity regarding a threat or construct.} 
The severity varies; for example, an article may have some paragraphs that are clear concerning both the threat and construct involved, whereas other paragraphs are generally unclear. Other articles discuss a threat in detail, but it is unclear which construct is associated with the threat or vice versa. For example, when an article states that threats to construct validity were mitigated because a discussion of the hypothesis of the experiment was avoided before it was conducted, it is unclear which constructs have avoided threats. Even though it is not explicitly stated what the threat is, we have coded this statement as "clear" regarding the threat because we easily understood that the threat considered was “hypothesis guessing”. Examples of unclear threats are descriptions of the purpose of an experiment or how a construct was measured without any threats indicated.

\begin{figure}[t]
  \centering
  \includegraphics[width=\linewidth]{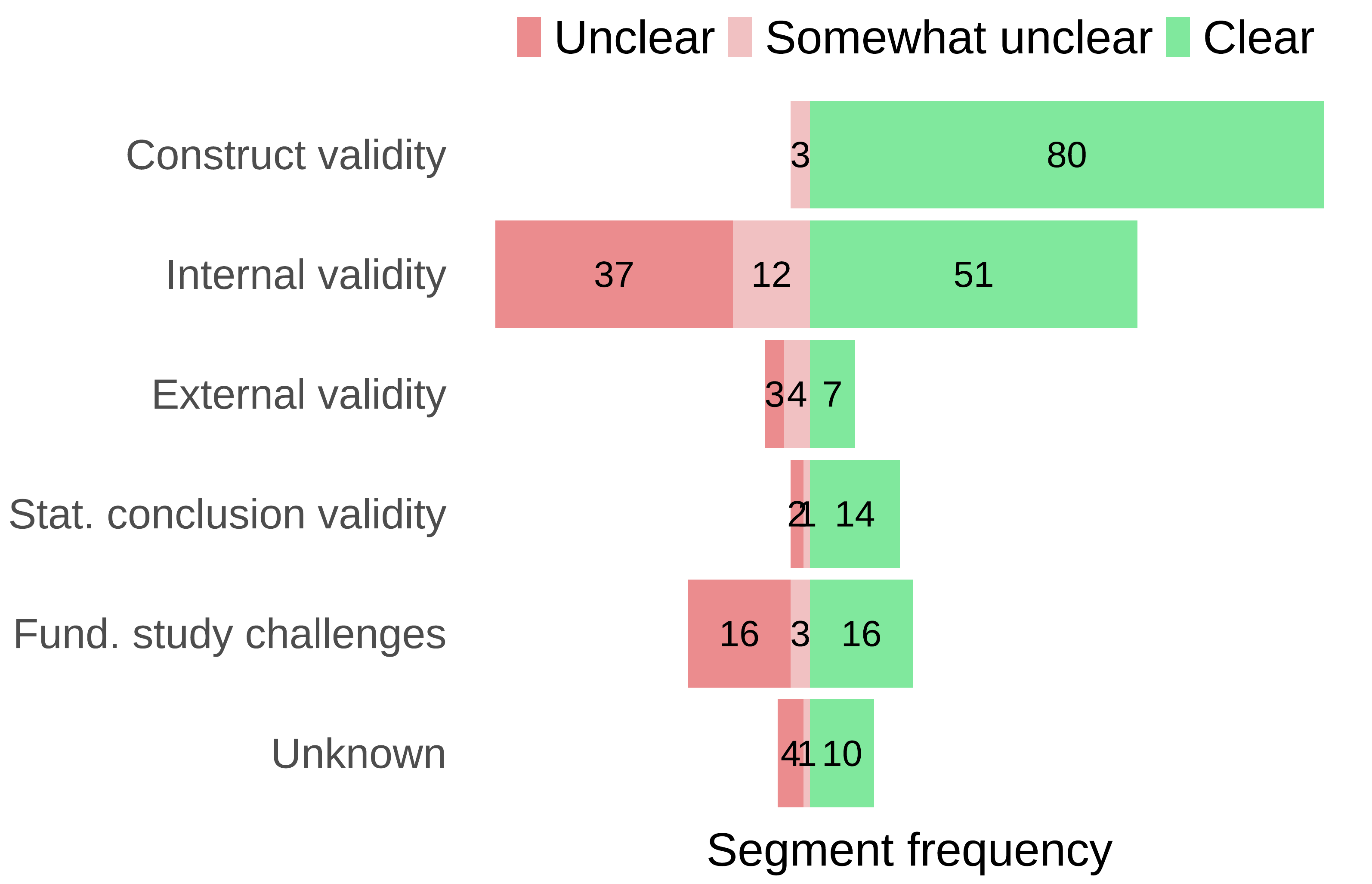}
  \caption{Clarity of constructs in segments by topic}
\end{figure}

Overall, the threats are unclear in 8\% and somewhat unclear in another 8\% of the segments. Segments that define CV or describe imported constructs are exempted from this analysis. Figure~3 shows that within a given topic, only statistical conclusion validity segments have a substantial proportion of unclear ($4/17$ segments = 24\%) or somewhat unclear (18\%) threats. An example of an unclear threat in this category is a segment that simply states that a five-point Likert scale was used. In this case, we could categorize the topic of the segment as statistical conclusion validity because the Likert scale was referred to. We manage to categorize one other unclear threat to fundamental study challenges. However, we cannot categorize the topic for most of the segments in which the threat is unclear. Consequently, most of the segments with unclear threats fall into the category of unknown. 

Constructs were more frequently unclear than threats; the constructs were unclear in 23\% and somewhat unclear in 9\% of the segments. Segments with definitions of CV are exempted from this analysis. Three-fifths of the articles have at least one segment with (somewhat) unclear constructs. Figure~4 shows that among the segments concerning CV, almost all constructs are clear (only three are somewhat unclear). However, for fundamental study challenges, there are as many unclear as clear constructs, and for internal validity, half of the segments have unclear or somewhat unclear constructs. 

Even though authors may believe that the threats and constructs in question are apparent from the context, they may not be apparent to readers. Therefore, we recommend that authors state threats and constructs clearly and explicitly. 
Note that if it is unclear which construct(s) is in question when a threat is described, it may not be a threat to CV but to other validity types.

\section{Limitations}
\label{sec:limitations}
A core limitation of our work is that we use a definition of CV on which not everybody may agree. 
We adhere to a minimal definition because we believe that such a definition is more 
coherent and, thus, comprehensible, which in turn is better for the research community.
Some of the threats and issues we consider outside the notion of CV may be covered by a broader definition of CV. 
For example, several of the segments that we have categorized as belonging to internal validity were considered as belonging to CV by Cook and Campbell in 1979~\cite{cook1979design}. Still, some of these types of threats, for example, "hypothesis guessing", were not considered CV threats any longer in 2002 by the same authors with the addition of  Shadish~et~al.~\cite{shadish2002experimental}. 
Shadish~et~al.~\cite{shadish2002experimental} discuss CV versus other validity types in further detail.

The identification of topics, CV definitions and their use in CV discussions, type of CV threats and clarity of threats and constructs is based on a subjective assessment by the authors working together and discussing until consensus is achieved. 
Although we did some automatic searches in the articles on a few specific keywords, for example, “mono-operationalization” and “mono-method”, the study basically faces the threat of mono-method; the single method was the authors’ subjective assessment. 
Instead of working together, we could have made the assessments independently and then 
reported agreement, for example, using inter-rater reliability. 

Regarding causal inferences, a plausible model would be that an unclear understanding of what construct validity entails by one or more authors results in a problematic article section on threats to construct validity. However, we have not investigated the causes of our findings. 

The sample we investigated comprised 83 articles that report human-centric experiments. 
We make no claims of generalization of the descriptive statistics beyond this sample. 
Consequently, external validity is not an issue. 
However, given that the articles were published in 
high-quality journals, one may conjecture that the challenges we report may be more prevalent in less prestigious publication venues.

\section{Conclusion}
\label{sec:conclusion}
The investigation of a sample of top-tier journal articles showed that threats to internal, external and statistical conclusion validity and problems with other fundamental study challenges were included under the umbrella of threats to CV. 
Furthermore, among those who introduced a definition of CV, the particular definition is not reflected in the discussion of CV for two-thirds of the articles. 
Finally, one-third of the analyzed text segments were unclear or somewhat unclear about which construct was involved in the asserted threat. 
For all the cases in which the construct was unclear, none of the threats were actually threats to CV. 

To improve the awareness and reporting of CV, we provided recommendations addressing the specific, identified weaknesses. Generally, we believe that CV and other validity types should be more prevalent in the literature and teaching of empirical research methods in software engineering. 


\bibliographystyle{ACM-Reference-Format}
\bibliography{bibliography}

\end{document}